# Reversible Image Authentication with Tamper Localization Based on Integer Wavelet Transform

P. Meenakshi Devi
Professor, Dept. of Information technology,
K.S.Rangasamy College of Technology,
Namakkal, Tamil Nadu, India
div_pri@yahoo.com

M. Venkatesan
Professor, Dept. of Master of Computer Applications,
K.S.Rangasamy College of Technology,
Namakkal, Tamil Nadu, India
venkatesh.muthusamy@gmail.com

K. Duraiswamy
Dean,
K.S.Rangasamy College of Technology,
Namakkal, Tamil Nadu, India

*Abstract*— In this paper, a new reversible image authentication technique with tamper localization based on watermarking in integer wavelet transform is proposed. If the image authenticity is verified, then the distortion due to embedding the watermark can be completely removed from the watermarked image. If the image is tampered, then the tampering positions can also be localized. Two layers of watermarking are used. The first layer embedded in spatial domain verifies authenticity and the second layer embedded in transform domain provides reversibility. This technique utilizes selective LSB embedding and histogram characteristics of the difference images of the wavelet coefficients and modifies pixel values slightly to embed the watermark. Experimental results demonstrate that the proposed scheme can detect any modifications of the watermarked image.

*Keywords: Reversible Watermarking; Authentication; Tamper Localization; Histogram modification;*

I. INTRODUCTION

Digital watermarking is the well known approach for image authentication reported in the literature. The traditional digital signature authentication methods have some drawbacks. As the signature is appended to a digital image, it increases the file size and it can also be removed easily. Moreover, it can not be used to locate the tampered area of an image with high accuracy. But, digital watermarking overcomes the above drawbacks.

In digital watermarking, the file size keeps unchanged after embedding the watermark also. Authentication watermark is very sensitive to any modifications imposed upon an image and can be used for tamper localization with high accuracy. In most conventional authentication techniques based on watermarking, the original image is distorted permanently due to the authentication itself. That is, the original image can not be recovered from the marked image when the marked image is deemed authentic. These distortions are not allowed in some sensitive applications, such as law enforcement, medical and military image systems. Thus, it is desired to undo the changes introduced by authentication if the image is verified as authentic. Data embedding techniques satisfying this requirement are referred to as *reversible or lossless* image authentication techniques. As long as the marked image is authentic, the original image can be reconstructed without any distortion. An effective authentication scheme basically should have the desirable features: tamper detection and localization, good perceptual invisibility, and detection without requiring explicit knowledge of the original image [1].

Many reversible authentication watermarking schemes for images have been reported in the literature [1]–[25], whereas rare of the schemes satisfy all the basic desirable features [2]. Most of the schemes do not provide tamper localization capability, which plays very important role in image authentication. Hierarchical authentication watermark in conjunction with the lossless generalized-LSB data embedding algorithm is used to offer localized lossless authentication watermark [22]-[25]. Hashes of IWT coefficients can also be used as watermark [3]. Some watermarking schemes apply data compression to reduce the size of embedding data [2]. The second category of reversible algorithms uses the difference expansion method to embed the watermark [4]-[9]. Integer Haar wavelet transform is applied to an image and the watermark is embedded into high-frequency coefficients by difference expansion [4]. Alattar extended Tian's scheme and applied the same difference expansion idea[5]. The third category of algorithms use histogram shifting method to embed the watermark [10]-[12]. In [10], the hash code of the image is combined with a binary logo image by a bit-wise exclusive OR and then embedded in the histogram of the difference image from the original image. Watermark is also embedded in wavelet coefficients in reversible manner [4], [12]-[15]. Lossless embedding of watermark is implemented with DCT coefficients [16]-[19]. Block-based reversible watermarking is implemented in [19].

In this paper, a new reversible authentication technique for images, which verifies the image integrity, identifies tamper detection, localization and reconstruction of original image is proposed. In order to verify the integrity of the image, a binary logo image is replicated to the size of the image and is embedded in selective LSB of the original image. The range of pixel value of the original image at the two extremes is narrow down to avoid underflow/overflow problems. To obtain reversibility, the original values of the modified LSBs and the pixels that are shifted to avoid underflow/overflow problems must also be sent as side information to the receiver as overhead data. IWT is applied on the watermark embedded image. A reversible watermarking technique using histogram modification is applied on the coefficients of high frequency sub-bands to embed the overhead data. Then inverse IWT is applied to get the final watermarked image.

## II. INVERTIBLE INTEGER-TO-INTEGER WAVELET TRANSFORMS

### A. Advantages of Integer Wavelet Transform

Conventional wavelet transform is not applicable to reversible authentication watermarking scheme since it does not guarantee the reversibility [113,114]. When an image is transformed into a wavelet domain using a conventional wavelet transform, the values of the wavelet coefficients will be the floating-point. If these coefficients are changed during the watermark embedding, the corresponding watermarked image block will not have accurate values. Any truncation of the floating-point values of the pixels may result in loss of information and may ultimately lead to the failure of the reversible authentication watermarking systems, that is, the original image can not be reconstructed from the watermarked image. Information may be lost through forward and inverse transforms. Furthermore, conventional wavelet transform is in practice implemented as a floating-point transform followed by a truncation or rounding since it is impossible to represent transform coefficients in their full accuracy. Hence, information is potentially lost through forward and inverse transforms [113]. To avoid this problem, an invertible integer-to-integer wavelet transform based on lifting is used in the proposed scheme. It maps integers to integers and does not cause any loss of information through forward and inverse transforms.

### B. Lifting Scheme Haar Transform

The wavelet Lifting Scheme is a method for decomposing wavelet transforms into a set of stages. Lifting scheme algorithms have the advantage that they do not require temporary arrays in the calculations steps and have fewer computations. When wavelet algorithms are expressed in a lifting scheme structure they are more efficient and they are simpler and easier to understand [115]. The lifting scheme consists of the following three steps:

- Split step – It is also called lazy wavelet transform. It divides the input data into odd and even elements.
- Predict step – It predicts the odd elements from the even elements.
- Update step – It replaces the even elements with an average.

The forward lifting scheme wavelet transform divides the data set being processed into an even half and an odd half. The index of $i^{th}$ element in the even half is represented as $even_i$ and the index of $i^{th}$ element in the odd half is represented as $odd_i$. A simple lifting scheme forward transform is depicted in Fig.1.

In the proposed scheme, lifting scheme version of the Haar transform is used. In this transform, the prediction step predicts that the odd element will be equal to the even element. The difference between the predicted value (the even element) and the actual value of the odd element replaces the odd element. For the forward transform iteration $j$ and element $i$, the new odd element, $j+1,i$ would be

$$odd_{j+1,i} = odd_{j,i} - even_{j,i} \quad (1)$$

In the lifting scheme version of the Haar transform the update step replaces an even element with the average of the even and odd pair.

$$even_{j+1,i} = \frac{even_{j,i} + odd_{j,i}}{2} \quad (2)$$

The original value of the $odd_{j,i}$ element has been replaced by the difference between this element and its even predecessor. The original value of $odd_{j,i}$ is recovered by using (3).

$$odd_{j,i} = even_{j,i} + odd_{j+1,i} \quad (3)$$

Substituting (3) into (2), $even_{j+1,i}$ becomes,

$$even_{j+1,i} = \frac{even_{j,i} + even_{j,i} + odd_{j+1,i}}{2}$$

$$even_{j+1,i} = even_{j,i} + \frac{odd_{j+1,i}}{2} \quad (4)$$

The averages (even elements) become the input for the next recursive step of the forward transform which is shown in Fig. 2.

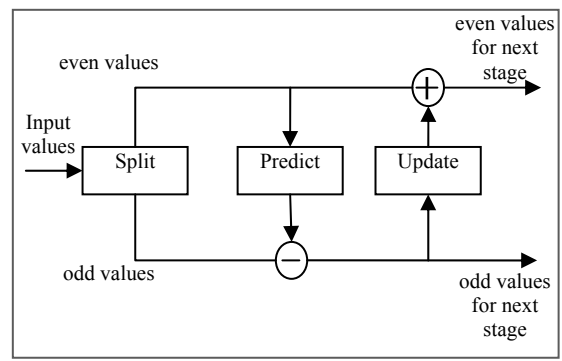

Figure 1 Lifting Scheme Forward Wavelet Transform

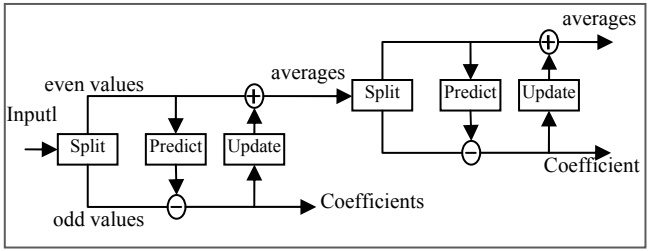

Figure 2 Two steps in the Lifting Scheme Forward Transform

One of the elegant features of the lifting scheme is that the inverse transform is a mirror of the forward transform. In the case of the Haar transform, additions are substituted for subtractions and subtractions for additions. The merge step replaces the split step. Fig. 3 depicts the lifting scheme inverse transform.

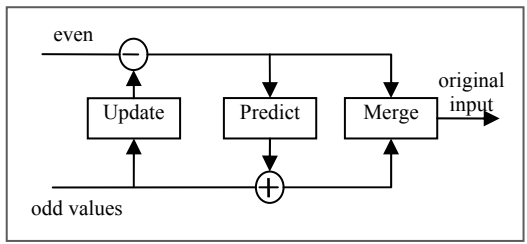

Figure 3 Lifting Scheme Inverse Transform

### III. PROPOSED REVERSIBLE AUTHENTICATION SCHEME

#### A. Pre-processing

When the watermark is embedded in the wavelet domain underflow or overflow can occur in the spatial domain. That is, the pixel values obtained from the watermarked wavelet coefficients can either be smaller than the minimum pixel value $p_{min}$ ($p_{min}$ = 0 for 8-bit gray-scale image) or be greater than the maximum value $p_{max}$ ($p_{max}$ = 255 for 8-bit gray-scale image). Since the reversibility is lost when underflow or overflow occurs, it must be predicted prior to the watermark embedding.

There are two ideas in the literature to avoid overflow/underflow problems. The first one identifies the pixels that cause overflow/underflow and ignores them during watermark embedding [8,9]. In this case, the payload must include information that provides the unused pixels in the watermark embedding process. The second one uses the concept of histogram shifting [19]. The pixels at the extreme levels are shifted towards the centre. In the proposed scheme, histogram shifting method is used to avoid overflow/underflow problems.

The range of pixel value of the original image is narrow down before applying IWT. Let *S* be shifting threshold, *x* and *x'* be pixel value before and after modification. For a 8-bit gray-scale image,

$$x' = x + S, if \ x \in [0, S] \quad (5)$$

$$x' = x - S, if \ x \in [255 - S, 255] \quad (6)$$

Now the range of pixel value is changed from [0,255] to [S, 255-S]. The pixels that are shifted must be recorded ( book keeping data ) and send to the receiver along with watermark as an overhead data.

#### B. Watermark Generation

The watermark to be embedded is taken as a either a logo or a random sequence of predefined bits which is known to both the sender and the receiver. The advantage of using the logo is that cropping can be easily detected.

The watermark is embedded in spatial domain, that is, LSB of selected pixels are modified to embed the watermark. The original image is divided into m × m sub-blocks (say, 3 × 3 or 5 × 5). The size of the sub-block defines the localization accuracy. If more accuracy is required, then the sub-block size should be minimum; otherwise, larger sub-blocks can be used. The watermark is generated by replicating the logo image so that the size of the watermark matches with the number of sub-blocks in the original image. Fig. 4 depicts the watermark generation process for lena image (256 × 256). The size of the sub-block is taken as 5 × 5 and hence the size of the logo is 51 × 51. The original logo and the generated watermark are shown in Fig. 4(b) and Fig. 4(c) respectively. To provide additional level of security, the generated watermark is scrambled using a shared secret key. Fig. 4(d) shows the scrambled watermark.

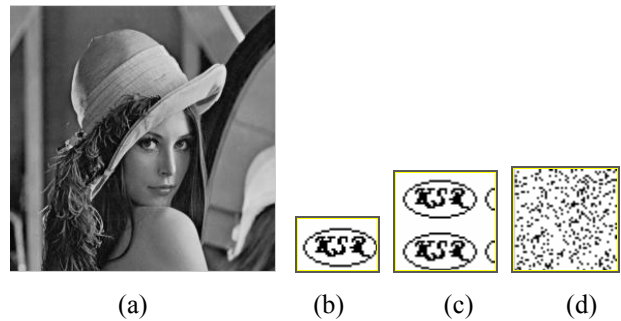

(a)  (b)  (c)  (d)

Figure 4 Watermark Generation for Lena Image (256 × 256). (a)Original Image; (b) original logo; (c) generated watermark and (d) Scrambled watermark

#### C. Watermark Embedding (Layer-1)

The original image *F* is divided into sub-blocks of size 5 × 5. Let the sub-block is represented as $F_{i,j}$, where *i, j* refers to the row and column of the sub-blocks. In each sub-block, one watermark bit is embedded. The centre pixel of the sub-block is modified to embed the watermark. The watermark bit is embedded as follows:

$$R_{i,j} = \mod\left(\sum_{a=1}^{m}\sum_{b=1}^{m}C_{a,b}, 2\right) \quad (7)$$

$$inc = R_{i,j} \oplus W_{i,j} \quad (8)$$

$$C_{centre} = C_{centre} + inc \quad (9)$$

$$lmap(i,j) = inc \quad (10)$$

where $C_{a,b}$ is the pixel value of the sub-block $F_{i,j}$, $R_{i,j}$ is the remainder in dividing the sum of pixel values of the sub-block $F_{i,j}$ by 2, $W_{i,j}$ is the watermark bit and $C_{centre}$ is the centre pixel of the sub-block $F_{i,j}$. If the reaminder $R_{i,j}$ and watermark bit $W_{i,j}$ are same, then *inc* will be zero and hence there is no need to make any changes in the sub-block; Otherwise, the centre bit, $C_{\lfloor m/2 \rfloor+1, \lfloor m/2 \rfloor+1}$ is incremented by *inc*(one). The result of the embedding process is Layer-1 Watermarked Image.

To obtain reversibility, the original values of the modified LSBs must also be sent as overhead data to the receiver. A '1' in the location map ($lmap(i,j)$) identifies the modified LSB. The location map is embedded in wavelet coefficients in reversible manner.

### D. Constructing Overhead Data

After embedding watermark in the LSBs of the image, the overhead data is embedded using reversible watermarking. It consists of *book keeping data* which provides information about the pixels shifted to avoid underflow/overflow and *location map* which provides information about the modified LSBs during watermark embedding process. This overhead data is embedded in wavelet domain in a reversible manner. The book keeping data is converted into a single bit stream as follows:

$$B = S \cup p \cup (r_1 \cup c_1) \cup (r_2 \cup c_2).....(r_p \cup c_p) \quad (11)$$

where *S* is the shifting threshold and *p* is the number of pixels shifted, $r_i$ and $c_i$ are the row and column of the $k^{th}$ pixel shifted. The size of location map is $M/5 \times N/5$. It is also represented as a single bit stream (*L*) by reading it in column wise. The overhead data bit stream *O* is formed by combining *B* and *L*.

$$O = B \cup L \quad (12)$$

The binary bit stream *O* can be compressed either using simple method like Run Length Encoding or using JBIG compression algorithm. The entire embedding process if depicted in Fig. 5.

### E. Overhead Data Embedding (Layer -2)

The L-1 watermarked image is decomposed by applying integer haar wavelet transform to get LL, HL, LH and HH sub-bands. The overhead data is embedded in high frequency sub-bands only. Let the size of the sub-bands are $X \times Y$. Difference images, $dD(i,j), dH(i,j), dV(i,j)$ of size $X \times Y/2$ are constructed for HH, HL and LH sub-bands respectively. For $1 \le i \le X$ and $1 \le j \le \frac{Y}{2}$,

$$\begin{aligned} dD(i,j) &= CD(i,2j) - CD(i,2j-1) \\ dH(i,j) &= CH(i,2j) - CH(i,2j-1) \\ dV(i,j) &= CV(i,2j) - CV(i,2j-1) \end{aligned} \quad (13)$$

where $CD(i,j)$, $CH(i,j)$ and $CV(i,j)$ are the coefficients of HH, HL and LH sub-bands respectively. The odd-line coefficients *(i, 2j-1)* are subtracted from the even-line coefficients *(i, 2j)* to construct the difference image.

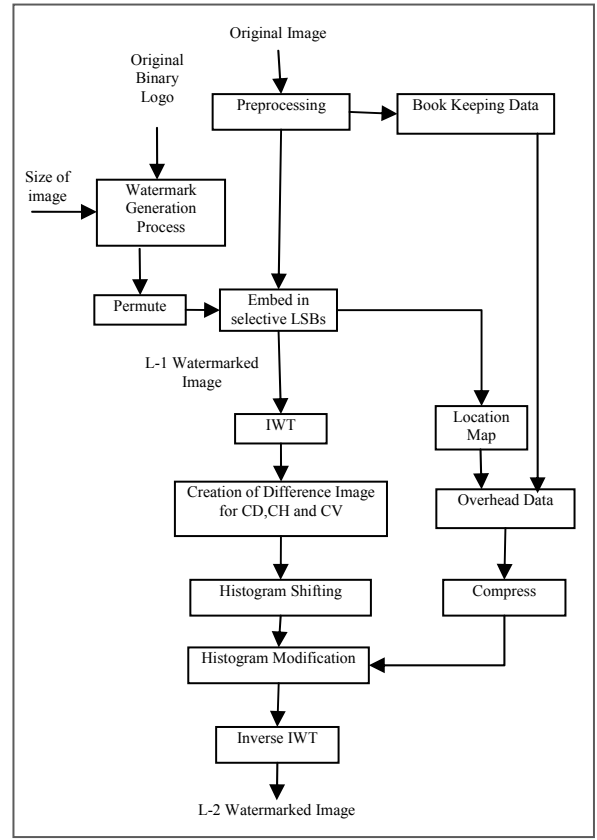

Figure 5 Watermark Embedding Process

The histogram shifting technique can be applied to make room for embedding overhead data [10-12]. The histogram bins of -2 and 2 are emptied by shifting some coefficient values in the difference images. If the difference value is greater than or equal to 2, then the even-line coefficient in the respective sub-band is incremented by one. If the difference value is less than or equal to -2, then the even-line coefficient in the respective sub-band is decremented by one. Then, the modified difference image is represented as $dD', dH'$ and $dV'$.

$$dD'(i,j) = CD'(i,2j) - CD'(i,2j-1)$$
$$dH'(i,j) = CH'(i,2j) - CH'(i,2j-1) \quad (14)$$
$$dV'(i,j) = CV'(i,2j) - CV'(i,2j-1)$$

Where,

$$CD'(i,2j) = \begin{cases} CD(i,2j)+1 & if\ dD(i,j \geq 2) \\ CD(i,2j)-1 & if\ dD(i,j \leq -2) \\ CD(i,2j) & otherwise \end{cases}$$

$$CH'(i,2j) = \begin{cases} CH(i,2j)+1 & if\ dH(i,j \geq 2) \\ CH(i,2j)-1 & if\ dH(i,j \leq -2) \\ CH(i,2j) & otherwise \end{cases}$$

$$CV'(i,2j) = \begin{cases} CV(i,2j)+1 & if\ dV(i,j \geq 2) \\ CV(i,2j)-1 & if\ dV(i,j \leq -2) \\ CV(i,2j) & otherwise \end{cases}$$

Now, the overhead data is embedded in sub-bands using the modified difference images. The order of difference images used is $dD', dH'$ and $dV'$. If $dD'$ and $dH'$ embeds the entire overhead information, then $dV'$ is not used. The modified difference images are scanned for embedding. Only the coefficients with the difference value of -1 and 1 are used. If such a value is encountered and if the overhead bit to be embedded is 1, then the difference value of -1 is made to -2 by subtracting one from the even-line coefficient or the difference value of 1 is made to 2 by adding one to the even-line coefficient. If the overhead bit is 0, the difference value of -1 or 1 is left unchanged. Based on the difference values, only the even-line fields of the coefficients are either incremented or decremented by one. The watermarked even-line coefficients of the three sub-bands are represented as follows: If the overhead data bit $O(k)=1$ and the difference value in the difference images is either 1 or -1, then the coefficients of the sub-bands are updated as follows:

$$CD_w(i,2j) = \begin{cases} CD'(i,2j)+1 & if\ dD'(i,j) = 1 \\ CD'(i,2j)-1 & if\ dD'(i,j) = -1 \end{cases}$$

$$CH_w(i,2j) = \begin{cases} CH'(i,2j)+1 & if\ dH'(i,j) = 1 \\ CH'(i,2j)-1 & if\ dH'(i,j) = -1 \end{cases} \quad (15)$$

$$CV_w(i,2j) = \begin{cases} CV'(i,2j)+1 & if\ dV'(i,j) = 1 \\ CV'(i,2j)+1 & if\ dV'(i,j) = -1 \end{cases}$$

The coefficients are left unchanged in all other cases.

$$CD_w(i,2j) = CD'(i,2j)$$
$$CH_w(i,2j) = CH'(i,2j) \quad (16)$$
$$CV_w(i,2j) = CV'(i,2j)$$

The odd-line fields of the watermarked sub-band coefficients are not affected in the embedding process and is given by,

$$CD_w(i,2j-1) = CD'(i,2j-1)$$
$$CH_w(i,2j-1) = CH'(i,2j-1) \quad (17)$$
$$CV_w(i,2j-1) = CV'(i,2j-1)$$

After the embedding process is over, inverse IWT is applied to obtain the L-2 Watermarked Image.

*F. Post processing*

The pixel values of the L-2 Watermarked Image are checked for overflow/underflow. If it does, the value of shifting threshold is incremented by one and the entire process is repeated once again; if it doesn't produce overflow/underflow, the watermarked image is ready to transmit.

*G. Watermark Extraction and Verification*

Fig. 6 depicts the watermark extraction and recovery scheme. At the receiving side, the first step is to extract the overhead data in the watermarked image.

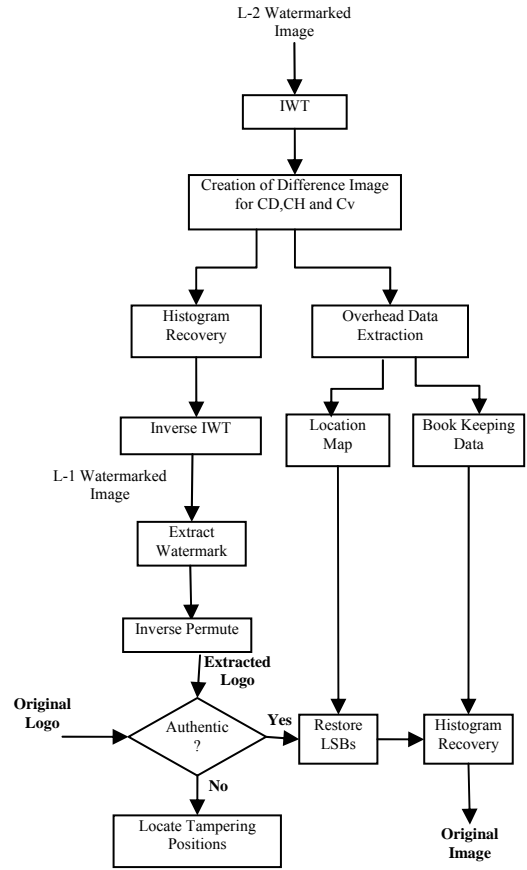

Figure 6 Watermark Extraction and Verification

IWT is applied on the watermarked image and the overhead data are extracted from the coefficients of high frequency sub-bands using reversible scheme. Then inverse IWT is applied to get the L-1 Watermarked Image. Now, the watermark bits embedded in the centre pixel of sub-blocks are extracted and the authenticity of the retrieved image is verified. If the image has not been tampered with and if the image is authentic, then the watermarked image is reversed back to its original form using the overhead data retrieved.

*H. Extraction of Overhead Data*

The difference images of the three high frequency sub-bands are found. The difference images are scanned in the same order as at the sender side. The overhead data is extracted as follows:

$$O_r(k) = \begin{cases} 0 & if\ dD_w(i,j) = 1\ or\ -1 \\ 1 & if\ dD_w(i,j) = 2\ or\ -2 \end{cases}$$

$$O_r(k) = \begin{cases} 0 & if\ dH_w(i,j) = 1\ or\ -1 \\ 1 & if\ dH_w(i,j) = 2\ or\ -2 \end{cases} \quad (18)$$

$$O_r(k) = \begin{cases} 0 & if\ dV_w(i,j) = 1\ or\ -1 \\ 1 & if\ dV_w(i,j) = 2\ or\ -2 \end{cases}$$

where, $O_r(k)$ is the overhead data retrieved, $dD_w$, $dH_w$ and $dV_w$ are the difference images of HH, HL, LH sub-bands of watermarked image respectively. It is not necessary to scan all the three sub-band coefficients. The size of the overhead data is also sent as a header in the payload. The two components of overhead data are separated as book keeping data and location map.

All the three difference images are scanned once again to recover the histogram shifting carried out during embedding process. Since only the even-line coefficients are manipulated, the odd-line coefficients of recovered image are directly obtained from the watermarked coefficients as follows:

$$CD_r(i, 2j-1) = CD_w(i, 2j-1)$$
$$CH_r(i, 2j-1) = CH_w(i, 2j-1) \quad (19)$$
$$CV_r(i, 2j-1) = CV_w(i, 2j-1)$$

Where $CD_w$, $CH_w$ and $CV_w$ are the coefficients of HH, HL, LH sub-bands of watermarked image and $CD_r$, $CH_r$ and $CV_r$ are the coefficients of HH, HL, LH sub-bands of recovered image. The even-line coefficients of the recovered image can be expressed as,

$$CD_r(i,2j) = \begin{cases} CD_w(i,2j)-1 & if\ dD_w(i,j) \geq 2 \\ CD_w(i,2j)+1 & if\ dD_w(i,j) \leq -2 \\ CD_w(i,2j) & otherwise \end{cases}$$

$$CH_r(i,2j) = \begin{cases} CH_w(i,2j)-1 & if\ dH_w(i,j) \geq 2 \\ CH_w(i,2j)+1 & if\ dH_w(i,j) \leq -2 \\ CH_w(i,2j) & otherwise \end{cases} \quad (20)$$

$$CV_r(i,2j) = \begin{cases} CV_w(i,2j)-1 & if\ dV_w(i,j) \geq 2 \\ CV_w(i,2j)+1 & if\ dV_w(i,j) \leq -2 \\ CV_w(i,2j) & otherwise \end{cases}$$

The inverse IWT is applied after the coefficients of recovered image are constructed. It is clear that, the resultant image is the L-1 Watermarked Image which contains only the watermark and the distortions due to overhead data embedding are removed. The next step is to verify the authenticity of the image.

*I. Image Authentication*

Since the watermark is embedded in LSBs of the centre pixel of $5 \times 5$ sub-blocks, it can be retrieved easily. The watermark bit embedded is retrieved as follows:

$$R_{i,j} = \mod\left(\sum_{a=1}^{m}\sum_{b=1}^{m} C'_{a,b}, 2\right) \quad (21)$$

$$W_{i,j} = R_{i,j} \quad (22)$$

where $C'_{a,b}$ is the pixel value of the sub-block $F'_{i,j}$, $R_{i,j}$ is the remainder in dividing the sum of pixel values of the sub-block $F'_{i,j}$ by 2, $W'_{i,j}$ is the watermark bit retrieved. The retrieved watermark is inverse permuted using the same secret key used at the sender side. The retrieved watermark is compared with the original watermark. If there are no distortions in the retrieved watermark, then the authenticity is verified; otherwise, it is concluded that the image might has been altered during transmission. Once the image authenticity is verified, the next step is to remove the distortions of watermark embedding and hence the recovery of the original image. If image authenticity is not verified, there is no need of recovery of original image; instead tamper localization is performed to identify the tamper positions.

*J. Recovery of Original image*

The overhead data retrieved at the earlier stage is used for recovering the original image. *Location map* provides information about the modified LSBs during watermark embedding process. A '1' in the location map ($lmap(i,j)$) identifies the modified LSB. It is used to recover the original values of LSBs. *Book keeping data* provides information about the pixels that are shifted to avoid underflow/overflow condition. The book keeping data is in a single bit stream format which is preceded by *shifting threshold (S)* and the number of pixels modified. Consecutive 16-bits are taken at a time in which the first 8-bits identify the row and the next 8-bits identify the column of the pixel shifted (for a 8-bit gray scale image).

Let $S$ be shifting threshold, $x'$ and $x$ be the pixel values of watermarked and the recovered original image. For a 8-bit gray-scale image,

$$x = x' - S, if\ x \in [0, S] \quad (23)$$
$$x = x' + S, if\ x \in [255-S, 255] \quad (24)$$

Now the range of pixel value is changed from [S, 255-S] to [0,255] and hence the original image is successfully recovered without any distortions.

## K. Tamper Localization

If authentication test is failed, it is clear that the watermarked image has been modified and tamper localization is applied to identify the tampering positions. Watermark is generated by replicating the watermark logo and permuted order is generated using the same shuffling key used at the sender side. It is compared with the watermark retrieved. If there is no match, then the respective $5 \times 5$ sub-block in the watermarked image is assumed to be modified. The following example illustrates the tamper localization clearly. Fig. 7(a) and 7(b) are the watermarked image and modified watermarked image respectively. The retrieved watermark is shown in Fig.7(c). It contains salt and pepper noise from which it is certain that the image has been tampered during transmission. The noise is extracted from the retrieved watermark and inverse shuffling is applied to locate the tamper positions in the original image. In Fig. 7(d), the white blocks depict the tamper positions. Table 1 lists the results of various images using the proposed scheme.

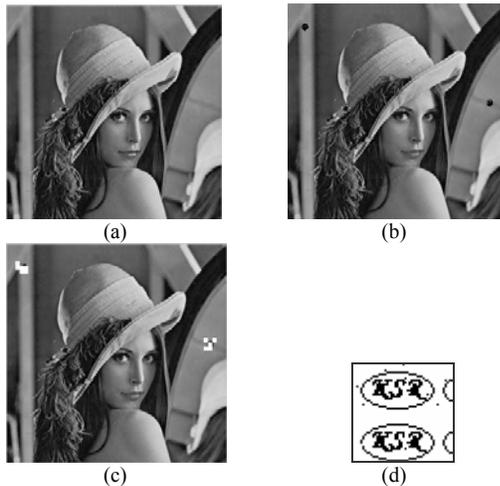

Figure 7. Tamper Localization; (a) Watermarked Image; (b) Tampered Watermarked Image; (c) Watermark Retrieved; (d) Image with tamper localization

TABLE I. EXPERIMENTAL RESULTS

| Image | Image Size in bits | PSNR in db | MSE |
|---|---|---|---|
| Lena | 256 × 256 | 51.4143 | 0.6852 |
| Baboon | 116 × 116 | 49.5756 | 0.8468 |
| Peppers | 136 × 137 | 50.3320 | 0.7761 |
| Barbara | 130 × 130 | 50.4714 | 0.7638 |
| Camera | 256 × 256 | 51.3475 | 0.6905 |
| Angel | 98 × 130 | 51.9675 | 0.6429 |
| Bridge | 131 × 90 | 50.4271 | 0.7677 |
| Dragon | 135 × 101 | 52.2837 | 0.6199 |

## IV. CONCLUSION

A wavelet-based reversible watermarking scheme for secure image authentication has been presented. In the proposed scheme, the embedded watermark is generated and scrambled based on the size of the image to be watermarked. This provides more protection to the watermarking system. Integer wavelet transform is applied and the proposed watermarking system localizes the tampering at pixel level. At the same time, if the image is deemed to be authentic, the original image can also be restored. Simulation results have been given to demonstrate the efficiency of the proposed scheme.

AUTHORS PROFILE

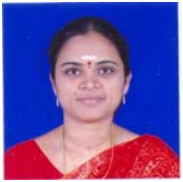
Ms. P.Meenakshi Devi received her B.E. degree in Kongu Engineering College, Perundurai, Tamil Nadu in 1993 and M.E. degree in Thiagarajar College of Engineering, Madurai, Tamil Nadu in 2003. She is a research scholar in the Department of Computer Science and Engineering. She is working as an Assistant Professor in Department of Information Technology, K.S.Rangasamy College of Technology, TamilNadu, India. Her area of interest includes Watermarking, Information Security, Cryptography and Computer Networks. She is a life member in ISTE.

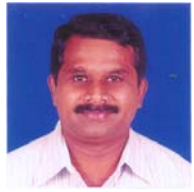
Mr. M.Venkatesan received his B.Sc.(Electronics) degree in Sri Ramakrishna Mission Vidyalaya Arts and Science College, Coimbatore, Tamil Nadu in 1990 and M.C.A. degree in Bharathidhasan University, Trichy, Tamil Nadu in 1997. He is a research scholar in the Department of Computer Science and Engineering. He is working as an Assistant Professor in Department of Master of Computer Applications, K.S.Rangasamy College of Technology, TamilNadu, India. His area of interest includes Watermarking, Information Security, Mobile Communication and Computer Networks. He is a life member in ISTE.

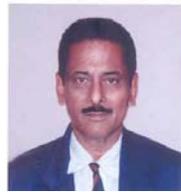
Dr. K.Duraiswamy (SM) received his B.E. degree in Electrical and Electronics Engineering from P.S.G. College of Technology, Coimbatore, Tamil Nadu in 1965 and M.Sc.(Engg) degree from P.S.G. College of Technology, Coimbatore, Tamil Nadu in 1968 and Ph.D. from Anna University, Chennai in 1986.

From 1965 to 1966 he was in Electricity Board. From 1968 to 1970 he was working in ACCET, Karaikudi, India. From 1970 to 1983, he was working in Government College of Engineering, Salem. From 1983 to 1995, he was with Government College of Technology, Coimbatore as Professor. From 1995 to 2005 he was working as Principal at K.S. Rangasamy College of Technology, Tiruchengode and presently he is serving as Dean in the same institution.

He is interested in Digital Image Processing, Computer Architecture and Compiler Design. He received 7 years Long Service Gold Medal for NCC. He is a life member in ISTE, Senior member in IEEE and a member of CSI.